\documentclass[manuscript,nolinenumbers]{aastex701}
\usepackage{bm}
\usepackage{amsmath}
\usepackage{color}
\usepackage{float}

\shorttitle{Tidal triggers and the 
predictability of solar activity}
\shortauthors{Stefani et al.}
\graphicspath{{./}{figures/}}

\begin{document}

\title{Tidal triggers and the 
predictability of solar activity}

\author[0000-0002-8770-4080]{Frank Stefani}
\affiliation{Institute of Fluid Dynamics, Helmholtz-Zentrum Dresden-Rossendorf, Bautzner Landstrasse 400, 01328 Dresden, Germany}
\email{F.Stefani@hzdr.de}

\author[0000-0001-9892-9309]{Gerrit M. Horstmann}
\affiliation{Institute of Fluid Dynamics, Helmholtz-Zentrum Dresden-Rossendorf, 
Bautzner Landstrasse 400, 01328 Dresden, Germany}
\email{G.Horstmann@hzdr.de}

\author[0000-0002-6189-850X]{George Mamatsashvili}
\affiliation{Institute of Fluid Dynamics, Helmholtz-Zentrum Dresden-Rossendorf, 
Bautzner Landstrasse 400, 01328 Dresden, Germany}
\affiliation{Abastumani Astrophysical Observatory, Mount Kanobili, Abastumani 0301, Georgia}
\email{G.Mamatsashvili@hzdr.de}

\author[0009-0007-0076-6531]{Tom Weier}
\affiliation{Institute of Fluid Dynamics, Helmholtz-Zentrum Dresden-Rossendorf, Bautzner Landstrasse 400, 01328 Dresden, Germany}
\email{T.Weier@hzdr.de}



\begin{abstract}
Magneto-Rossby waves in the solar tachocline are currently considered to be one 
of the main determinants of solar activity.
In particular, they can give rise to the 
quasi-biennial 
oscillation (QBO). The latter was recently shown to be dominated by a 
phase-stable period of around 1.7 years. 
By analyzing 72 ground-level enhancement (GLE) events and 37 S-flares, we 
determine that this period is close to 1.723 years. This, in turn, is 
the dominant beat between the 
periods of the spring tides of the tidally dominant planets Venus, Earth, 
and Jupiter, which are suspected to synchronize not only the 
QBO, but 
also the 11.07-year Schwabe cycle.  We demonstrate that recent events, such as the solar 
storm of 2024 May 10 and the strong X-flare  of 2026 February 1, align 
well with maxima of the combined tidal forcing. 
\end{abstract}

\keywords{Solar physics; Solar dynamo; Tidal Interaction }

\section{Introduction} \label{sec:intro}
The importance of Rossby waves for terrestrial weather systems and their 
predictability has been recognized for almost a century 
\citep{Rossby1939}. However, Rossby waves may also play a 
key role in space weather which is governed by solar activity \citep{Zaqarashvili2010a,Zaqarashvili2010,Dikpati2012,Marquez-Artavia2017,Dikpati2018,Zaqarashvili2021}. 
As first discussed by \cite{Zaqarashvili2010}, they can 
naturally explain the Sun's  quasi-biennial oscillations (QBO).
Using the shallow-water magnetohydrodynamic approximation, these authors 
demonstrated that the interaction between differential rotation and a 
toroidal magnetic field exceeding $10^5$\,G leads to the instability 
of magneto-Rossby waves with a period of approximately two years. 
Later on, \cite{Raphaldini2015}  argued that the dynamics of a 
resonant triad of magneto-Rossby waves could lead to periodically 
changing wave amplitudes with periods comparable to the 
dominant 11-year Schwabe cycle.

In addition to the huge energy reservoirs provided by differential rotation 
and toroidal magnetic fields, there are other sources that can 
cause neutrally stable Rossby waves to become unstable.
One of them was recently discussed by \cite{Horstmann2023} who 
considered the tidal forces exerted by the revolving planets 
on the Sun. When forced by realistic-amplitude 
tides, magneto-Rossby waves were shown to acquire velocities 
of the order of m/s or larger, 
depending on a damping parameter whose precise value is, however,  
still unknown.
Based on this result, it was demonstrated by \cite{Stefani2024} that 
the two-planet spring tides of Venus-Jupiter (with period 118 days), 
Earth-Jupiter (199 days), and Venus-Earth (292 days) lead to a 
beat period of 11.07 years which corresponds remarkably well to the Schwabe cycle.
Saturn, whose tidal effect on the Sun is much weaker, only comes 
into play when considering the rosette-like barycentric motion of the 
Sun, which is known to be governed by the 19.86-year Jupiter-Saturn 
synodic cycle. 
While the internal flow structure triggered by the interaction 
of this barycentric motion with the Sun's rotation around its 7$^{\circ}$ 
inclined axis is still under scrutiny \citep{Shirley2023,Pierron2026}, 
its inclusion in our dynamo model via a parametrization produced a long-term 
solar activity spectrum that remarkably aligns with that observed 
in climate-related sediments from Lake Lisan (see Figure 9 in 
\cite{Stefani2024}).
The most striking feature of the emerging spectral peaks is the 193-year 
beat period, which appears to correspond to the Suess-de Vries cycle.
 
Yet another, much shorter beat period of 1.723\,years was revealed by 
\cite{Stefani2025} and shown to coincide precisely with the dominant period 
underlying the occurrence of ground level enhancement (GLE) events. 
This  work also confirmed the remarkable finding 
of \cite{Velasco2018} that those GLE events exhibit phase 
stability over nearly six solar cycles.

In view of this phase stability, and the time lag of 
approximately 1.7 years between the 
solar storm of 2024 May 10 and the strong
X-class flare of 2026 February 1, 
it is tempting to take a closer look into 
the potential predictability of space weather 
events\footnote{When asked by spaceweather.com about the potential 
impact of the upcoming Venus-Earth-Jupiter alignment 
on solar activity, one of us (F.S.) made on 2026 January 7 the 
following prediction: ``If the alignment excites magneto-Rossby 
waves as our model predicts, we might expect a higher probability 
of strong solar activity 40 to 60 days from now'' 
(https://spaceweather.com/archive.php?view=1\&day=07\&month=01\&year=2026). 
Unfortunately, due to a bug in the underlying code, this estimate was 
off by one month and should have read ``10 to 30 days from now'', which 
would have included the 2026 February 1
event. Further 
details on that issue will be discussed below.}.
This is precisely the subject matter of the present paper. 
It goes beyond our previous work \citep{Stefani2025} in two respects. 
Firstly, in addition to the
72 GLE events, we will analyze a series of 37 S-class 
events for which we will also find phase stability with a 
dominant period that is very close to that of the GLE events.

Secondly, we will compute the correlations of a
merger of GLE and S-flare data not only with
a single cosine function, but also with the
actual beat signal resulting from the
superposition of the three planetary tidal triggers. 
We will see that a significant portion of the events occur very 
close to the peaks of that beat signal.
This particularly applies to the solar events 
of 2024 May 10 and 2026 February 1.

The paper will close with some conclusions and 
an outlook on further work.

\section{Ground level enhancement events}
In this section we revisit the GLE 
events that were analyzed previously by \cite{Velasco2018} and
\cite{Stefani2025}. These sporadic events are related to 
relativistic solar particles that produce air showers, 
the effects of which can be measured at ground level 
by a network of detectors. Table 1 and Figure 3 
of \cite{Velasco2018} revealed that the 56 considered 
GLE events occurred preferentially
in the positive phase of an oscillation with a 
period of 1.73\,years which indeed points to a
clocked process that was phase stable over 
approximately six solar cycles.

\begin{table}[]
\scriptsize
\caption{Numbers, dates in the format YYYY/MM/DD, 
and days elapsed since 1955/12/31 of 
the 72 GLE events considered here. Note the beginning 
of the numbering with 5, and the missing event 17, 
according to the dataset provided at 
https://gle.oulu.fi .}
\begin{tabular}{ccccccccc}
\hline
\hline
No & Date & Days& No & Date & Days & No & Date & Days \\
\hline
5  & 1956/02/23 & 54    & 30 & 1977/11/22 & 7997  & 54 & 1992/11/02 & 13456 \\
6  & 1956/08/30 & 243   & 31 & 1978/05/07 & 8163  & 55 & 1997/11/06 & 15286 \\
7  & 1959/07/17 & 1294  & 32 & 1978/09/23 & 8302  & 56 & 1998/05/02 & 15463 \\
8  & 1960/05/04 & 1586  & 33 & 1979/08/21 & 8634  & 57 & 1998/05/06 & 15467 \\
9  & 1960/09/03 & 1708  & 34 & 1981/04/10 & 9232  & 58 & 1998/08/24 & 15577 \\
10 & 1960/11/12 & 1778  & 35 & 1981/05/10 & 9262  & 59 & 2000/07/14 & 16267 \\
11 & 1960/11/15 & 1781  & 36 & 1981/10/12 & 9417  & 60 & 2001/04/15 & 16542 \\
12 & 1960/11/20 & 1786  & 37 & 1982/11/26 & 9827  & 61 & 2001/04/18 & 16545 \\
13 & 1961/07/18 & 2026  & 38 & 1982/12/08 & 9839  & 62 & 2001/11/04 & 16745 \\
14 & 1961/07/20 & 2028  & 39 & 1984/02/16 & 10274 & 63 & 2001/12/26 & 16797 \\
15 & 1966/07/06 & 3840  & 40 & 1989/07/25 & 12260 & 64 & 2002/08/24 & 17038 \\
16 & 1967/01/28 & 4046  & 41 & 1989/08/16 & 12282 & 65 & 2003/10/28 & 17468 \\
18 & 1968/09/29 & 4656  & 42 & 1989/09/29 & 12326 & 66 & 2003/10/29 & 17469 \\
19 & 1968/11/18 & 4706  & 43 & 1989/10/19 & 12346 & 67 & 2003/11/02 & 17473 \\
20 & 1969/02/25 & 4805  & 44 & 1989/10/22 & 12349 & 68 & 2005/01/17 & 17915 \\
21 & 1969/03/30 & 4838  & 45 & 1989/10/24 & 12351 & 69 & 2005/01/20 & 17918 \\
22 & 1971/01/24 & 5503  & 46 & 1989/11/15 & 12373 & 70 & 2006/12/13 & 18610 \\
23 & 1971/09/01 & 5723  & 47 & 1990/05/21 & 12560 & 71 & 2012/05/17 & 20592 \\
24 & 1972/08/04 & 6061  & 48 & 1990/05/24 & 12563 & 72 & 2017/09/10 & 22534 \\
25 & 1972/08/07 & 6064  & 49 & 1990/05/26 & 12565 & 73 & 2021/10/28 & 24043 \\
26 & 1973/04/29 & 6329  & 50 & 1990/05/28 & 12567 & 74 & 2024/05/11 & 24969 \\
27 & 1976/04/30 & 7426  & 51 & 1991/06/11 & 12946 & 75 & 2024/06/08 & 24997 \\
28 & 1977/09/19 & 7933  & 52 & 1991/06/15 & 12950 & 76 & 2024/11/21 & 25163 \\
29 & 1977/09/24 & 7938  & 53 & 1992/06/25 & 13326 & 77 & 2025/11/11 & 25518 \\
\hline
\end{tabular}
\end{table}

\begin{figure}[t]
  \centering
\includegraphics[width=0.85\textwidth]{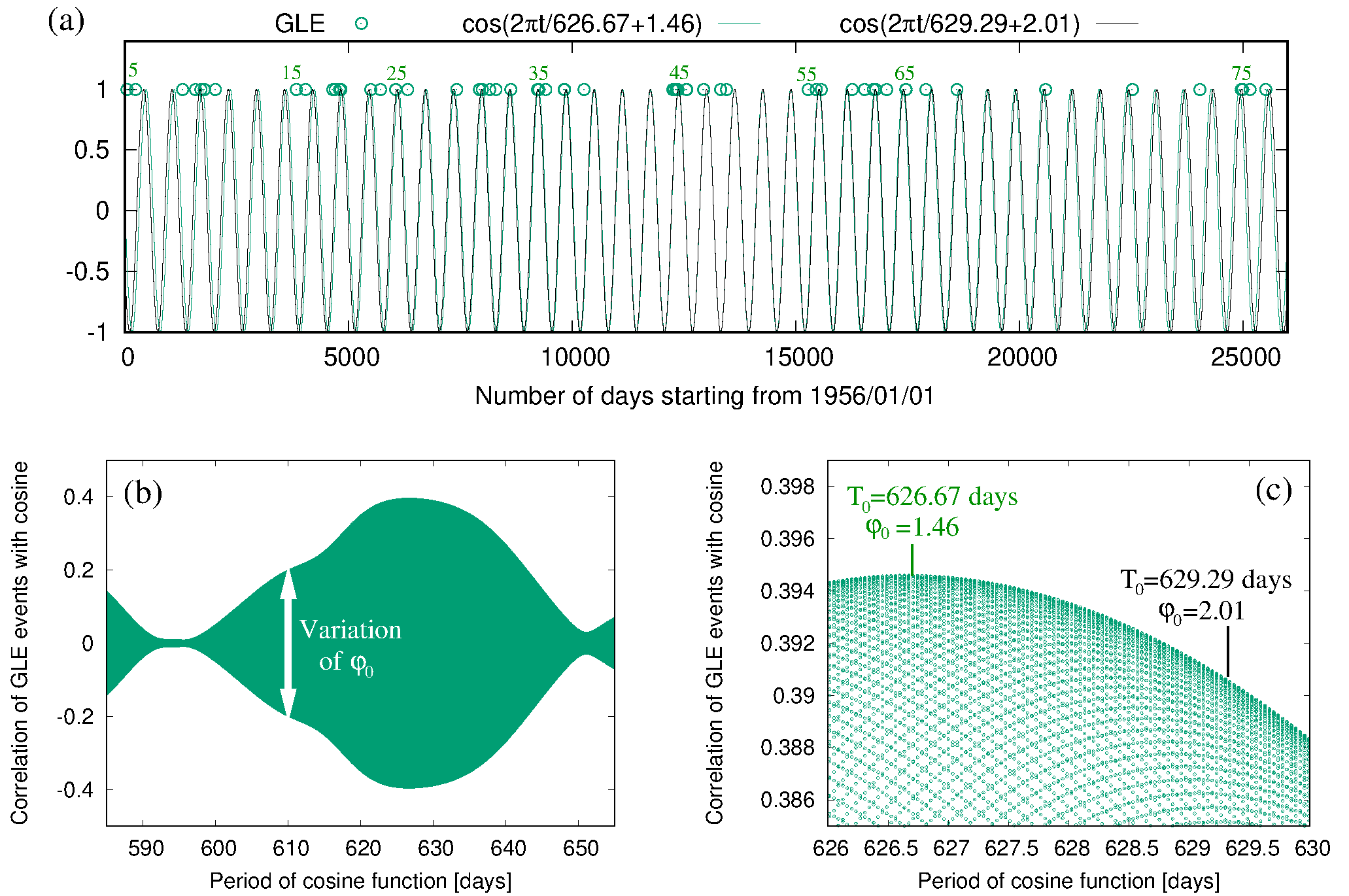}
  \caption{Analysis of GLE data.
  (a) Distribution of 72 events 
  (green open circles) observed between 1956 February and 
  2025 November, as obtained from https://gle.oulu.fi. 
  The abscissa shows the time $t$ in 
  days starting from 1956 January 1.
  A few event numbers referring to Table 1 are indicated.
  The green curve shows the optimum cosine 
  function $\cos(2 \pi t/626.67 + 1.46)$
  that maximizes the correlation coefficient given in Equation (2).
  The black curve displays 
  $\cos(2 \pi t/629.29 + 2.01)$ 
  that results from the phase optimization
  when the theoretical 
  period of $T_0=629.29$ days is fixed beforehand. 
  (b) Correlation 
  coefficient of the 72 GLE events with 
  cosine functions with variable 
  periods $T_0$ and phases
  $\varphi_0$. For each period $T_0$, the
  vertical extent emerges when varying  
  $\varphi_0$ between 0 and $2 \pi$.
  (c) Zoomed-in version of (b), showing the maximum of $\rm Corr$ appearing for $T_0=626.67$\,days and 
  $\varphi_0=1.46$. If the period is fixed beforehand to 
  $T_0=629.29$ days, we obtain a slightly decreased correlation coefficient for 
  an optimum $\varphi_0=2.01$. These periods and phases are used for 
  defining the green and black curves in (a).}
\end{figure}

In \cite{Stefani2025} we had re-analyzed these GLE data, updating them from the 56 events used in 
\cite{Velasco2018} to the 71 events provided
by the database of neutron monitor count rates at https://gle.oulu.fi.
To precisely determine the best-fit period, 
we replaced the inverse wavelet method of  \cite{Velasco2018} by computing 
the correlation coefficient 
\begin{eqnarray}
r=1/N \sum_{i=1}^{N} \cos(2 \pi t_i/T_0+ \varphi_0)
\end{eqnarray} 
of 
$N=71$ GLE instants $t_i$, using
cosine functions with variable 
periods $T_0$ and phases
$\varphi_0$. While $r$ is not exactly Pearson's 
empirical correlation coefficient, it shares with it the 
main property of 
lying between -1 and 1. 
The latter value only occurs when 
the events and the cosine's maxima are perfectly aligned.

Figure 8 of \cite{Stefani2025} 
showed a maximum correlation of 0.2964 for a period of
629.85 days, which corresponds to 1.724 years. 
This value is indeed remarkably close to the value of 
1.723 years we had derived as the 
dominant beat of the 
tidally triggered magneto-Rossby waves.

In the following, two modifications will be applied. Firstly, we correct a timing 
error in \cite{Stefani2025} 
that led to an incorrect phase of $\varphi_0=1.88$, 
when 
it should have been $\varphi_0=2.19$.
While this phase error
(equivalent to a time shift of 31 days) had no 
effect on the main message of \cite{Stefani2025}, 
it will be important later on when considering the 
predictability of the solar events of early 2026.

Secondly, from here on we will use 
a modified measure of 
correlation in the form 
\begin{eqnarray}
{\rm Corr}=\frac{ \sum_{i=1}^{N} \cos(2 \pi t_i/T_0+ \varphi_0)}
{\sqrt{\sum_{i=1}^{N} \cos^2(2 \pi t_i/T_0+ \varphi_0)}}
\end{eqnarray}
which is closer in spirit to  
Pearson's correlation coefficient between 
events (characterized by a number 1) and
the underlying cosine function.  We select this metric 
having in mind its straightforward transferability to more 
complex functions.
Yet, in Appendix A, we will revisit the correlation $r$, as 
defined in Equation (1),
when employing the Rayleigh test, which 
is the standard method for detecting periodicity in 
unevenly distributed data \citep{Droege1990}.

Table 1 lists the 72 GLE events between February 1956 and November 2025.
The results of the data analysis are shown in  
Figure 1. Panel (a) displays the distribution of the 
72 GLE events over time. Panels (b) and (c) illustrate the 
determination of the optimal period $T_0=626.67$\,days and 
phase angle $\varphi_0=1.46$, which are  used to 
define the green curve in panel (a).
Additionally,
we also determine the optimum phase when 
fixing $T_0$ beforehand to the theoretical value
of 629.29 days as derived from the Rossby-wave theory. 
The resulting cosine function is displayed in black in 
panel (a). 
The two curves are almost indistinguishable 
from each other, with a mere 0.4 per cent difference in period.
This suggests that the underlying theory may indeed
have some merit. Note also that the two correlation values obtained, 
0.394 and 0.391, correspond to two-sided p-values of 0.00062 and 
0.00068 for a sample size of 72, which indicates a high level of significance for the correlation.
Admittedly, since $\rm Corr$ is not exactly a Pearson
correlation coefficient, those $p$-values should be
interpreted with some caution.

\section{S-class flares}

In this section we apply the methodology outlined in the
previous section to the 37 events of solar superflares 
of S-class ($>$X10 in soft X-rays) 
that were recorded since 1978 \citep{Tan2025}. 
The occurrence of 
these S-flares 
was recently shown in the wavelet analysis of
\cite{Velasco2026} 
to be governed by coupled phase states of 1.7-year and 7-year oscillations.
The dates of these events, 
taken from \cite{Tan2025} and \cite{Velasco2026}, are  listed in Table 2.

\begin{table}[]
\scriptsize
\caption{Number, dates in the format YYYY/MM/DD, 
and days elapsed since 1955/12/31 of the 37 S-flare events 
considered here. The data are from \cite{Tan2025} and \cite{Velasco2026}.}
\begin{tabular}{ccccccccc}
\hline
\hline
No & Date & Days   & No & Date & Days & No & Date & Days \\
\hline
1 & 1978/7/11 & 8228 & 14 & 1990/5/24 & 12563 & 27 & 2001/4/15 & 16542 \\
2 & 1980/11/6 & 9077 & 15 & 1991/1/25 & 12809 & 28 & 2003/10/28 & 17468 \\
3 & 1982/6/3 & 9651 & 16 & 1991/3/4 & 12847 & 29 & 2003/10/29 & 17469 \\
4 & 1982/6/6 & 9654 & 17 & 1991/3/22 & 12865 & 30 & 2003/11/2 & 17473 \\
5 & 1982/7/12 & 9690 & 18 & 1991/6/1 & 12936 & 31 & 2003/11/4 & 17475 \\
6 & 1982/12/15 & 9846 & 19 & 1991/6/4 & 12939 & 32 & 2005/1/20 & 17918 \\
7 & 1982/12/17 & 9848 & 20 & 1991/6/6 & 12941 & 33 & 2005/9/7 & 18148 \\
8 & 1984/4/24 & 10342 & 21 & 1991/6/9 & 12944 & 34 & 2006/12/5 & 18602 \\
9 & 1984/5/20 & 10368 & 22 & 1991/6/11 & 12946 & 35 & 2011/8/9 & 20310 \\
10 & 1989/3/6 & 12119 & 23 & 1991/6/15 & 12950 & 36 & 2017/9/6 & 22530 \\
11 & 1989/8/16 & 12282 & 24 & 1992/11/2 & 13456 & 37 & 2017/9/10 & 22534 \\
12 & 1989/9/29 & 12326 & 25 & 1997/11/6 & 15286 & - & - & - \\
13 & 1989/10/19 & 12346 & 26 & 2001/4/2 & 16529 & - & - & - \\
\hline
\end{tabular}
\end{table}

\begin{figure}[t]
  \centering
\includegraphics[width=0.85\textwidth]{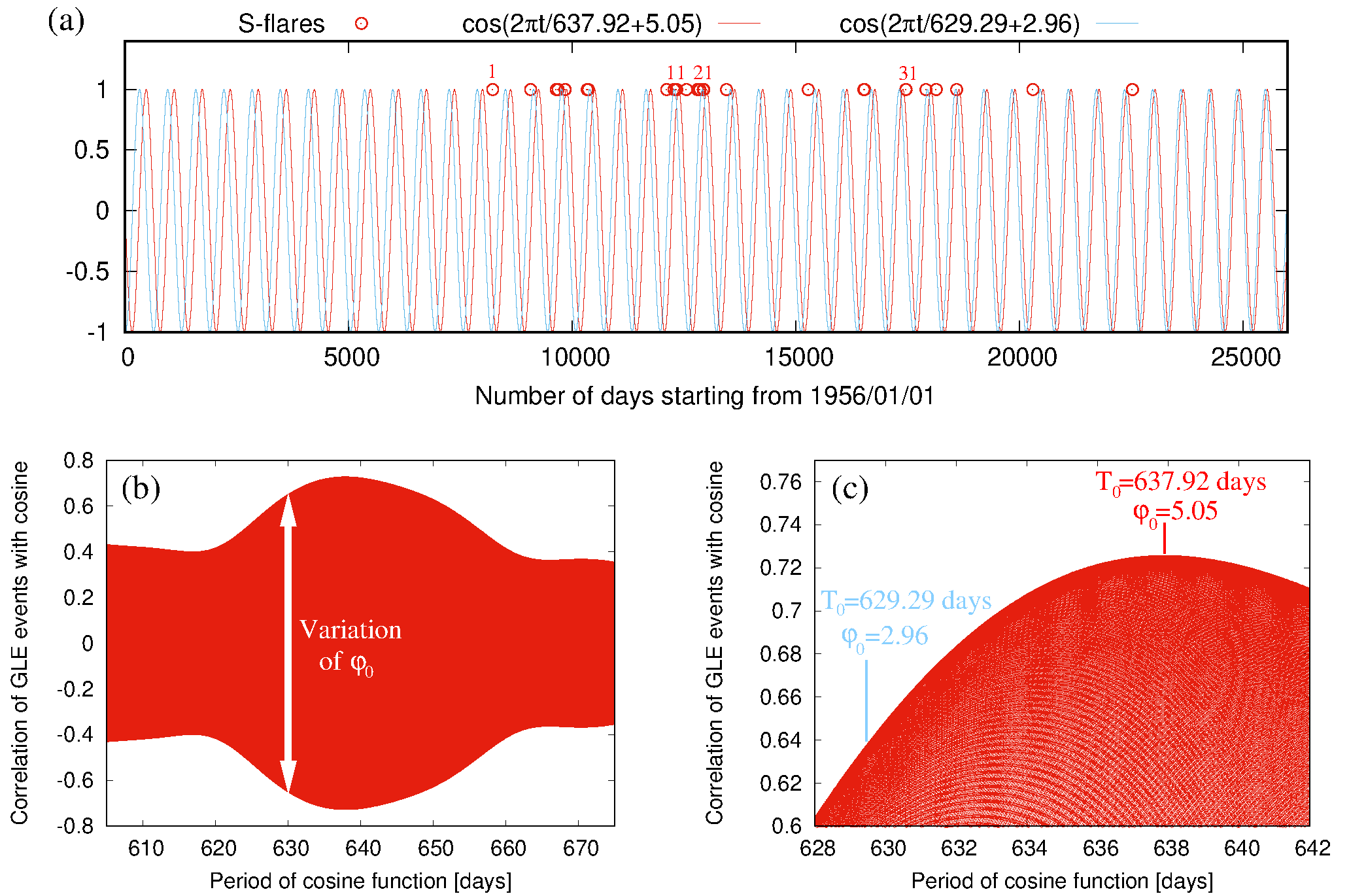}
  \caption{Same as Figure 1, but for the 
  37 S-flare events since 1978. Here, the red curve shows the optimum cosine 
  function $\cos(2 \pi t/637.92 + 5.05)$
  that maximizes $\rm Corr$ given in Equation (2).
  The light blue curve shows 
  $\cos(2 \pi t/629.29 + 2.96)$ 
  that results from the optimization
  of the phase when the theoretical 
  period of $T_0=629.29$ days is fixed beforehand. 
  }
\end{figure}

The data analysis is illustrated in Figure 2. Panel (a) shows 
the distribution of the 
37 S-flare events, while in panels (b) and (c) 
the optimal periods $T_0$ and 
phase angles $\varphi_0$ are determined. 
Here, the best-fit period, $T_0=637.92$ days, is a bit 
farther away 
(1.4 per cent) from the theoretical value of 629.29 days.
For $N=37$, the two correlations of 0.72 for the optimum
period of 1.747 years and 0.64 for the theoretical period 
of 1.723 years have p-values as low as $5.5\times 10^{-7}$ 
and $3.6 \times 10^{-5}$, respectively,  which implies an even 
higher significance of the correlations
than in the GLE case.

\section{GLE and S-flare events taken together}
Given the very close results obtained for the best-fit 
periods of the 72 GLE events and the 37 S-flares, we 
decided to assess also a combined dataset of 109 events 
in total. This is illustrated in Figure 3. 
Here, the resulting best-fit period, $T_0=632.88$ days, is 
only 0.6 per cent away from the theoretical value of 629.29 days.
For $N=109$, the two correlations of 0.435 for the optimum
period of 1.733 years and 0.424 for the theoretical period 
of 1.723 years have p-values of 
$3.2\times 10^{-6}$ and $6.7\times 10^{-6}$, 
respectively.

\begin{figure}[t]
  \centering
\includegraphics[width=0.85\textwidth]{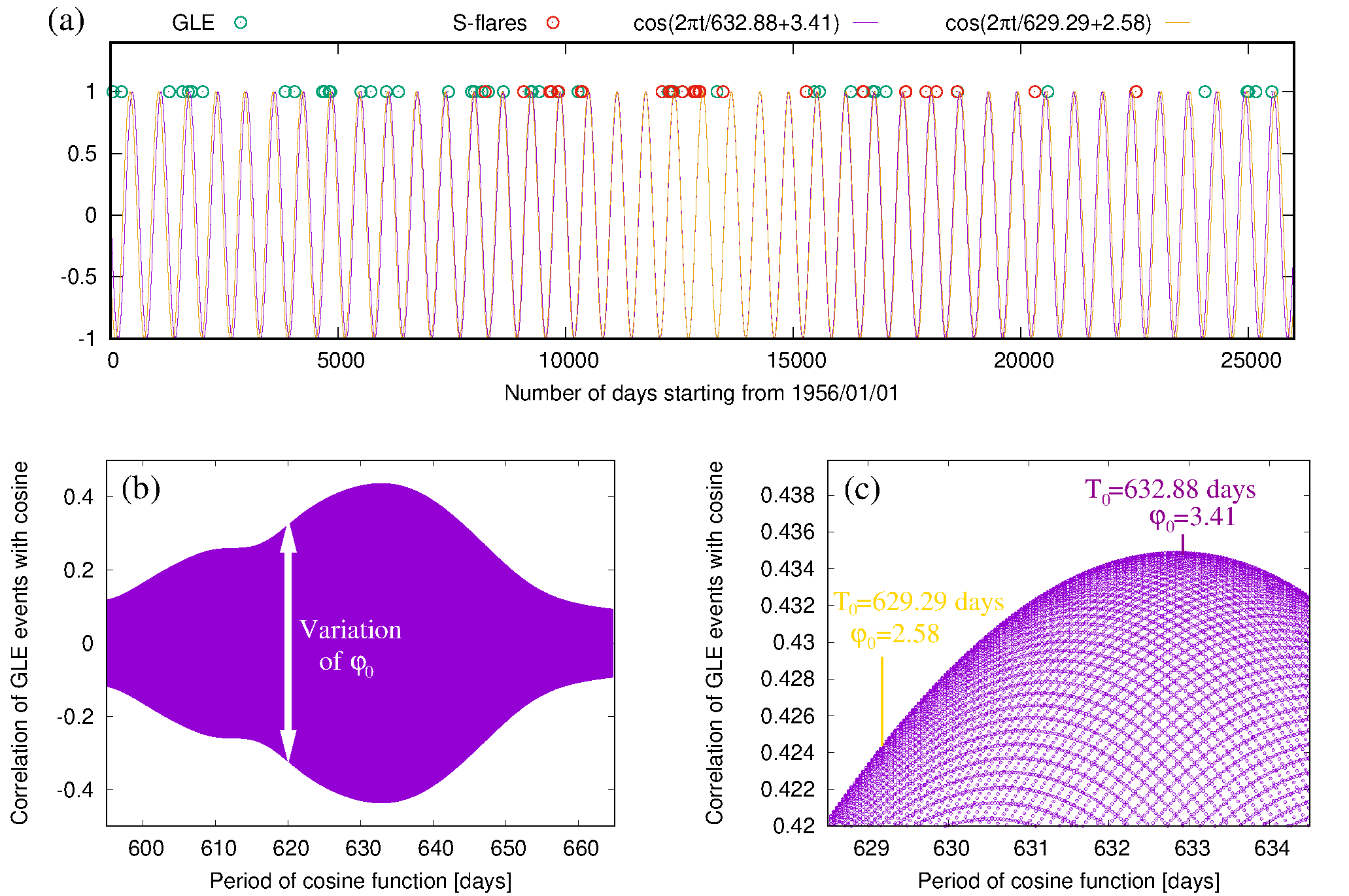}
  \caption{Same as Figure 1, but for the merger of the
  72 GLE (green open circles) with the 
  37 S-flare events (red open circles). Here, 
  the violet curve shows the 
  optimum cosine 
  function $\cos(2 \pi t/632.88 + 3.41)$.
  The gold curve shows 
  $\cos(2 \pi t/629.29 + 2.58)$ 
  that results from the optimization
  of the phase when the theoretical 
  period of $T_0=629.29$ days is fixed beforehand.}
\end{figure}

While all previous analyses used simple
cosines as test functions, in the following 
we move closer to the Rossby-wave mechanism
underlying the synchronization theory of the solar dynamo
that we have developed over the last decade
\citep{Weber2015,Stefani2016,Stefani2019,Stefani2021,Klevs2023,Horstmann2023,Stefani2024,Stefani2025}, 
building on previous work of 
\cite{Abreu2012,Scafetta2012,Wilson2013,Okhlopkov2016}.

Following \cite{Stefani2025}, we first consider
the sum of three equally weighted tidal wave excitations 
with periods of (approximately) 118 days, 199 days and 292 days that correspond 
to the two-planet spring-tides of Venus-Jupiter, Earth-Jupiter and Venus-Earth,
respectively:
\begin{eqnarray}
s(t)&=& \cos\left( 2\pi  \cdot \frac{t-t_{\rm VJ}}{0.5 \cdot P_{\rm VJ}}\right) +\cos\left( 2\pi  \cdot \frac{t-t_{\rm EJ}}{0.5 \cdot P_{\rm EJ}} \right)+\cos\left( 2\pi  \cdot \frac{t-t_{\rm VE}}{0.5 \cdot P_{\rm VE}} \right) \;.
\end{eqnarray}
To be more specific, we use 
the accurate two-planet 
synodic periods $P_{\rm VJ}=0.64884$\,years, 
$P_{\rm EJ}=1.09207$\,years,
$P_{\rm VE}=1.59876$\,years, and the 
epochs of the corresponding conjunctions
$t_{\rm VJ}=2002.34$,
$t_{\rm EJ}=2003.09$, and
$t_{\rm VE}=2002.83$
that were adopted from 
\cite{Scafetta2022}.

As shown by \cite{Dikpati2020a}, 
magneto-Rossby waves lead to tachocline bulges and depressions 
in the top surface and can therefore play a 
role in providing longitude and latitude 
locations for 
emergence of toroidal fields into the 
convection zone and photosphere.
Moreover, the maximum field-strength of toroidal flux 
tubes that can be stored in mechanical equilibrium 
at the bottom of the convection zone is very sensitive 
to the value of the superadiabaticity
\citep{Abreu2012}.
In this sense, the sum $s(t)$ of the three tidal forcings 
might already be relevant for the excitation of
tachocline deformations and the launching of flux tubes.
However, in the following we will consider its square, $s^2(t)$, which may be
more relevant for tachocline nonlinear oscillations (TNO),
modifications of zonal or meridional flows, or 
the wave-related helicity and the dynamo-relevant 
$\alpha$-effect.

The square $s^2(t)$ of the sum of the three cosine functions 
is shown as the dark blue curve 
in panels (a) and (b) of Figure 4, together
with the 72 GLE and 37 S-flare events. 
Compared to Figures 1, 2, 3,
we have doubled the time-resolution 
in order to better recognize the 
frequent (although not perfect) coincidences 
of solar events with the  spikes of $s^2(t)$.
While the average distance between these spikes corresponds to 
the above-mentioned beat period of 1.723 years (629.29 days), 
they show a systematic oscillation
leading to the additional 11.07-year beat period
which can be recognized in the changing height of the spikes.
As an aside, this second period is at the root of our 
parametric resonance model to describe the
Schwabe cycle
\citep{Stefani2019,Klevs2023,Stefani2024,Stefani2025}.

In the following we will assess the correlation 
of the tidal forcing $s^2(t)$
with the solar events which we define, in analogy to Equation (2), as 
\begin{eqnarray}
{\rm Corr}&=&\frac{ \sum_{i=1}^{N} [s^2(t_i)-\overline{s^2(t)}]   }
{\sqrt{\sum_{i=1}^{N}  [s^2(t_i)-\overline{s^2(t)}]^2}} \; .
\end{eqnarray}
In this context, it seems also indicated to 
``smear out'' the very spiky shape of
$s^2(t)$ by applying to it a moving average.
Indeed, the toroidal magnetic field, on which the 
amplitudes of the three tidally excited magneto-Rossby 
waves depend, exhibits a complicated butterfly-type variation in space and time.
While finding the most appropriate averaging procedure over 
space and/or time is non-trivial, we consider averaging over a 
small fraction of the Schwabe cycle to be reasonable.
Three of those representative averages (with time-windows of 101, 201, and 301 
days) are displayed 
in panels (a) and (b) of Figure 4 with different colors.
For the original $s^2(t)$ and the three time-averages  we ask 
now which time-shift
(backward or forward) would give the highest correlation
with the solar event data.
While formally this corresponds to the  
phase optimization for the single cosine functions
as carried out in Figures 1, 2, and 3, 
now the time-shift has a clear physical meaning in that it 
describes the lag between the maximum tidal triggering of 
Rossby waves and the solar event that ultimately 
occurs at the solar surface. 
In view of the typical rise times of flux tubes in the order of a month or more 
\citep{Weber2011}, one would expect a similar time lag to come out as a sort of optimum.

However, the picture arising from panels (c) and (d) of Figure 4 is less clear. 
While in general the correlation peaks around zero time lag, 
in particular the dark blue 
curve of $s^2(t)$ has a  somewhat 
counterintuitive peak at a 
{\it positive} time shift of approximately 
70 days with respect 
to the solar events. 
For longer time averages (yellow and gold), 
the correlation curves become much smoother 
and indeed peak around a zero time lag.

What is remarkable in this respect is the 
case of an average window of 101 days. 
For this
window length the main peaks in panels (a) and (b) 
come to lie 
in between the peaks of the original 
signal $s^2(t)$.

In Appendix B, we will assess whether the correlation 
can still be improved by varying the pre-factors of 
the three individual waves, which entered Equation 
(3) with equal weights.

\begin{figure}[t]
  \centering
\includegraphics[width=0.87\textwidth]{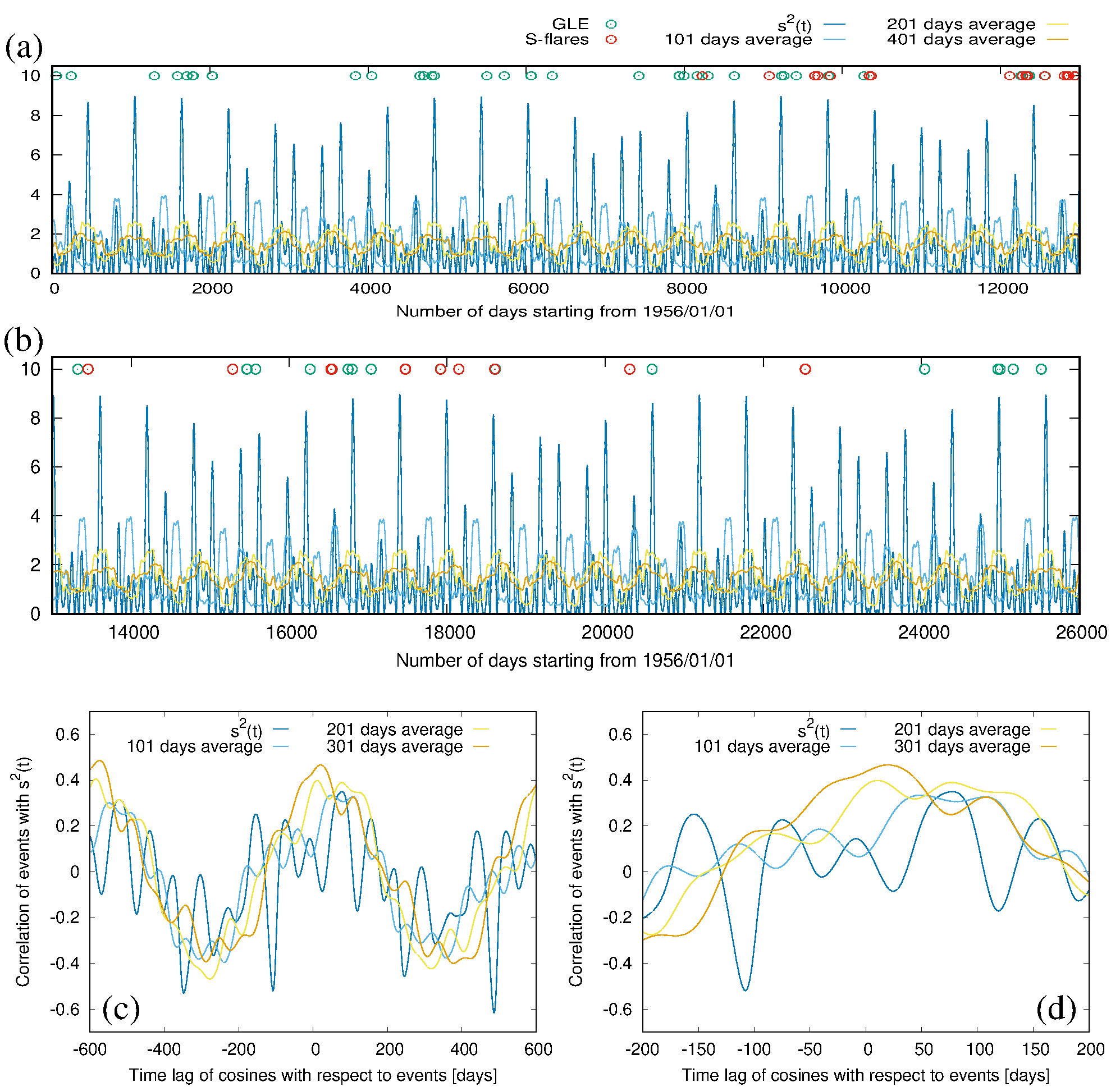}
  \caption{GLE and S-flare events, $s^2(t)$, 
  and different time averages of it (a,b). The longer beat 
  period of 11.07\,years (corresponding to the Schwabe cycle) 
  can be recognized in the varying
  height of the spikes. Correlation coefficients of the 
  four curves 
  from (a,b) 
  with the solar events in dependence on the time-lag (c). 
  The same, but zoomed in on shorter time lags (d).}
\end{figure}

\section{X-flare of 2026 February 1}

The high solar activities of early 2026 
occurred
approximately 1.7 years after the two 
GLE events of 2024 May and June.
In this section, we will assess the 
possible connection between these events and the strong 
tidal forcings resulting from the alignment of Venus, 
Earth and Jupiter.

To start with, we display in Figure 5(a) the X8.3-class flare
of 2026 February 1 (red full triangle)
in the context of the solar events since late 1977 November 
(8000 days after 1955/12/31). We also plot  $s^2(t)$, 
alongside the black, light blue and gold cosine curves from 
Figures 1, 2 and 3, which had come out as the best fits 
when the period was fixed beforehand to 629.29\,days.
Obviously, all these curves are very close to 
each other, and the spiky $s^2(t)$ curve fits 
nicely into the picture, despite its slight wiggling 
to the left and right associated with the occurrence of 
double peaks. Overall, we also observe a significant, 
albeit not perfect, synchronism between the peaks 
of the four curves and the solar events.
 
The focus in panel (b) of Figure 5 is now on the last two years.  
On the left-hand side, we recognize the two 
GLE events that occurred on 2024 May 11 
and 2024 June 8.
Remarkably, the peak of $s^2(t)$ lies almost exactly 
between these two closely neighboring events.
Here, the idea that solar events are not related to the 
very maximum of the $s^2(t)$ function, but to the neighbouring points of its steepest ascent 
or descent, comes to mind.
Very close to $s^2(t)$ lies the much wider black
curve, representing the best-fit cosine function
of the GLE events for the fixed 629.29-day period.
The other two cosine curves (the light blue and gold ones) are shifted 
slightly towards earlier times.
Turning to the right-hand side of panel (b), we 
see that the strong X-flare event of 2026 February 1 occurred only 
23 days after the peak of $s^2(t)$ located  on 2026 January 9. While this 
corresponds nicely to the point of 
steepest {\it descent} of $s^2(t)$, the preceding GLE 
event of 2025 October 25 occurred a bit too early to coincide with the point of 
steepest {\it ascent}.

The dates of the maxima of the pure cosines are also displayed in panel (b).
Note that the sequence of the maxima of the 
dark blue and the black curve 
has interchanged between 2024 and 2026. While 
both have the same dominant 
period of 629.29 days, 
$s^2(t)$ has slightly wiggled to the left.

In panel (c) we show the corresponding  
green, red and violet single-cosine curves 
from Figures 1, 2, 3 which were obtained when 
optimizing both the phases {\it and} the periods.
Once again, we see a fairly good correlation with 
the solar events.

We note in passing that omitting the 
one GLE-event from 2025 in the determination of the 
optimal parameters leads only
to a minor shift of the corresponding curves
in Figure 5 by 3-16 days to the right.

In this context, let us finally take a look at the dashed and 
dotted green curves in panel (c).
The dotted one, $\cos(2 \pi t/629.85 + 1.88)$, was 
presented in \cite{Stefani2025} as the best fit when 
using the simplified correlation expression
$r=1/71 \sum_{i=1}^{71} \cos(2 \pi t_i/T_0+ \phi_0)$. 
However, this solution was flawed due to a bug in the code, and 
it should have been the dashed green curve, i.e.  
 $\cos(2 \pi t/629.77 + 2.19)$.
The 31-day difference between them explains the inaccuracy of the forecast 
made on 7 January 2026, as mentioned above. In any case, in view of the results obtained in the present paper, we would generally argue in favour of using $s^2(t)$ instead of the cosine functions for any forecasting purposes.
Indeed, both the two GLE events of 2024 and
the strong X-flare of 2026 February 1
occurred very close to the 
peaks of $s^2(t)$.
The timing of the latter event aligns closely with the expected one-month lag for the flux tubes to emerge on the Sun's surface \citep{Weber2011}, although this may be a coincidental occurrence. As mentioned above,
it could also be the points of steepest ascent or descent that are decisive.

\begin{figure}[t]
  \centering
\includegraphics[width=0.7\textwidth]{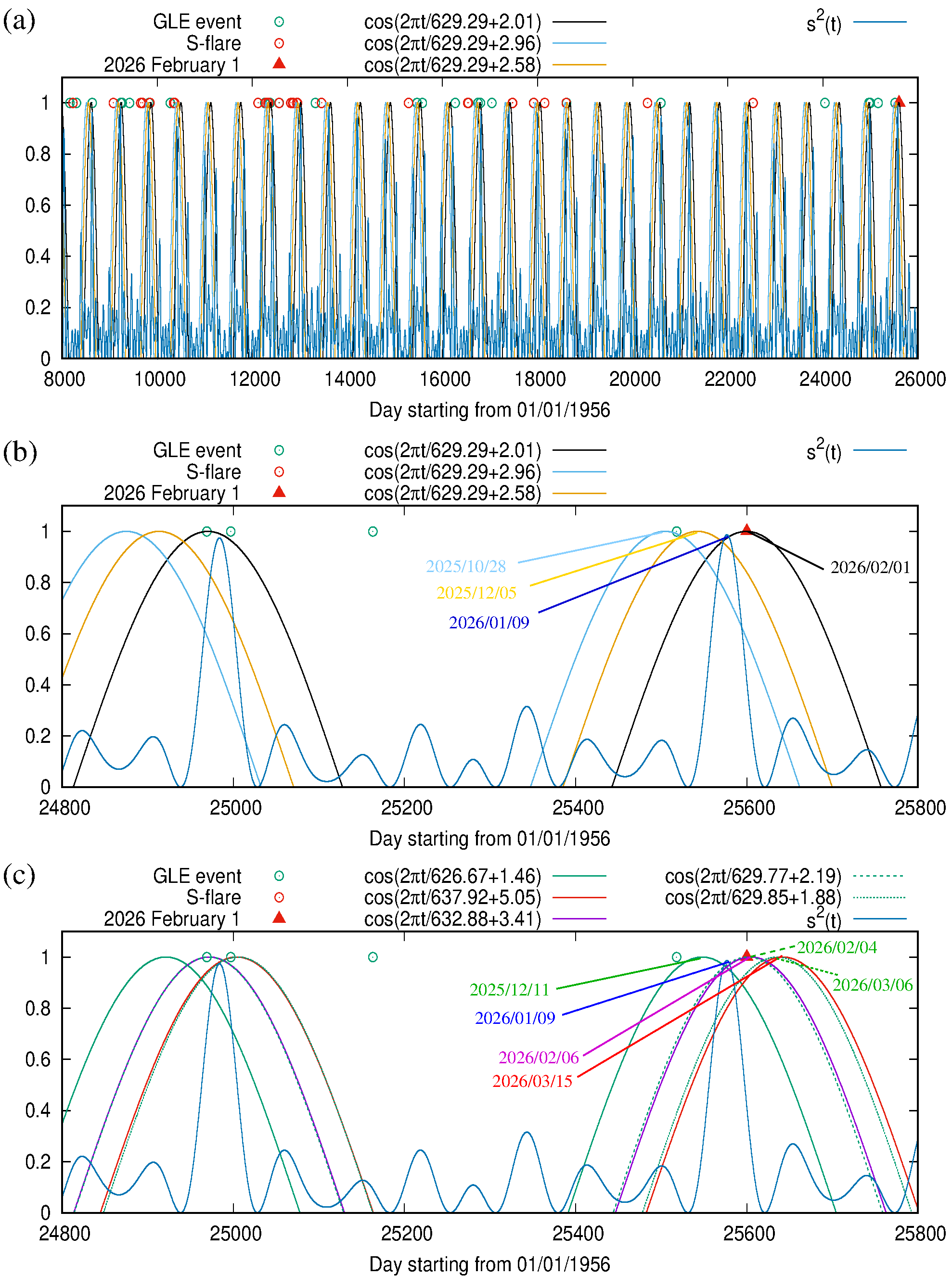}
  \caption{The 2026 February 1 event and its potential predictability.
  (a) GLE and S-flare events since 1977 November together with
  $s^2(t)$ and the three optimized 629.29-day periodic functions. The 
  strong X-flare of 2026 February 1
  is added
  as a red triangle. 
  (b) Details of (a) with the
  dates of the maxima of the four curves indicated. (c) 
  Modification of (b) showing the curves with optimized phases $\varphi_0$ and 
  periods $T_0$. The dashed green curve  with $\varphi_0=2.19$ would
  result when using 
  the  correlation function without denominator. The dotted 
  green line, with the 
  erroneous phase $\varphi_0=1.88$, 
  was given in 
  \cite{Stefani2025}. 
  The phase difference corresponds to the time shift of 31 day by 
  which the prediction for the early 2026 events was off.}
\end{figure}

\section{Conclusions}
In this paper we have refined and deepened 
our previous examination of a potential link between 
extreme solar events and tidal triggers of 
magneto-Rossby waves at the solar tachocline.
As for the GLE events, we found a correlation 
coefficient of 0.394 when using a cosine function with 
an optimized period of 1.716 years. Utilizing the 
theoretical beat period of 1.723 years gave a 
slightly reduced correlation of 0.391. 
For the 72 GLE events used, the corresponding 
low p-values pointed to a high significance of the
correlations.

When applying the same method to the sequence of 
37 S-class flares, we found even higher correlations of 0.72 
and 0.64 for the optimized and theoretical periods of 1.747 
and 1.723 years, respectively. The 
corresponding $p$-values were very low.

We also applied the method to a simple merger of the GLE and S-flare events.
For the arising 109 events we obtained 
correlations of 0.434 for an optimum period
of 1.733 years and 0.424 for 1.723 years, 
leading again to rather low $p$-values of 
$3.2\times 10^{-6}$ and $6.7\times 10^{-6}$, 
respectively.

As shown in Appendix A, the use of the Rayleigh test 
instead of the correlation coefficient $\rm Corr$ 
leads only to minor modifications of our results. 

Optimizing the functions with fixed periods for different sets 
of solar events results in a maximum phase difference of 15 per 
cent, equivalent to 95 days.
This, along with the highly significant correlations, indicates 
a phase-stable process that governs the solar QBO.

For the merged data set, we also computed 
the correlation of the events with the
square of the sum of the three tidal trigger functions which is 
dominated by 
steep peaks separated, on average, by gaps of 629.29 days (1.723 years).
When shifting this function (and its various moving averages) 
backwards or forwards in relation to the solar events, we found that the maximum correlation 
typically occurred with a positive time lag 
of around 70 days. At first glance, this is surprising, since 
one would expect the reverse, meaning that tidal peaks 
should precede solar events.
One possible explanation is the significant width 
(of approximately 100 days) of the main peaks of the $s^2(t)$ curve, 
which could lead to flux tube launches already at the rising 
flank of the tidal trigger of the Rossby waves.

All this shows that we still have a long way to go before we 
understand the physical mechanism that links the excitation 
of magneto-Rossby waves by tides with the launching of flux tubes. 
It goes without saying that any viable forecast must also take 
into account the actual phase of the 11-year Schwabe cycle.
In this context, the third harmonic of the Hale cycle may also play a role, 
as suggested by recent results of \cite{Velasco2026} and confirmed in 
Appendix A. Apart from this, we must also acknowledge that although 
magneto-Rossby waves provide an ideal resonance ground for tidal 
forces to act upon, it cannot be ruled out that other mechanisms are at play.

In the future, we will  assess how much further the correlations can be 
improved by taking into account the dependence of the weights of the three 
waves on the instantaneous toroidal field. It may also be worthwhile
including somewhat less extreme events, 
such as X-flares and sub-GLE events, in the data analysis.

\section*{Appendix A}

In Sections 2-4 we have considered the correlation coefficient 
$\rm Corr$, defined in Equation (2), 
as an appropriate metric that can be easily generalized 
to other 
test functions. In this Appendix, we apply the Rayleigh test 
to the datasets as a complementary method of establishing a 
connection with more standard techniques for determining 
dominant periods from non-evenly distributed data.

\begin{figure}[h]
  \centering
\includegraphics[width=0.95\textwidth]{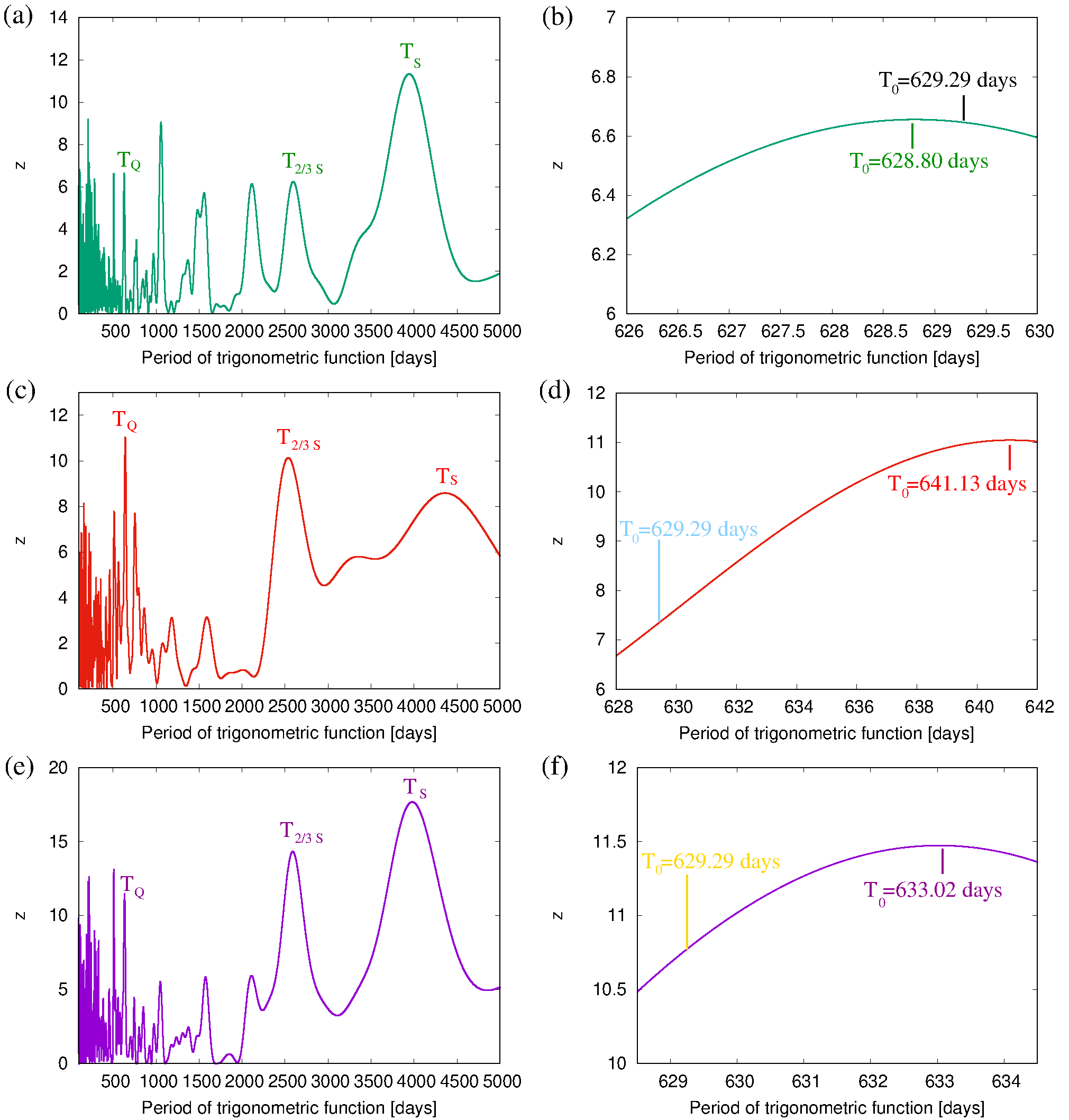}
  \caption{Results of the Rayleigh tests for the 72 GLE events (a,b),
  the 37 S-flare events (c,d), and the 109 merged events (e,f).
  The left column (a,c,e) shows the $z$-values between 100 and 5000 days,
  the right column shows the $z$-values in the vicinity of 629.29 days.
  Here, the chosen segment of the abscissa corresponds to those
  of the (c)-panels in Figures 1, 2 and 3.
  The specific $z$-values at the optimized $T_0$ are, 
  respectively, 6.66 (a), 11.04 (b), and 11.47 (c). The corresponding 
  values  at the theoretical QBO-period are,
  respectively, 6.65 (a), 7.27 (b), and 10.78 (c).}
\end{figure}

According to \cite{Droege1990}, the test statistic $z$ 
used in the Rayleigh test is
defined by
\begin{eqnarray}
z(T_0)&=&\frac{1}{N} \left[ \left( \sum^N_{i=1} \cos(2 \pi t_i/T_0) \right)^2+\left( \sum^N_{i=1} \sin(2 \pi t_i/T_0) \right)^2 \right] \, .
\end{eqnarray}
Again, $t_i$ stands for the instants of the $N$ extreme solar events 
under consideration, and $T_0$ is the variable period.
Note that the term in the square brackets corresponds to 
the square of the measure $r$, defined previously 
in  Equation (1), when
using therein the optimized phase $\varphi_0$ that 
maximizes the sum of the cosine-terms, while 
zeroing the sum of the sine-terms.
This lends plausibility to the notion $z=N r^2$ which is 
frequently used in
in the context of the Rayleigh test.

Figure 6 shows the  $z$-values obtained when using the 72 GLE events (a,c),
the 37 S-flare events (b,d), and the 109 merged events (e,f).
Panels (a,c,e) on the left-hand side  show the general picture of 
$z(T_0)$ between 
100 and 5000 days. In addition to the obvious QBO-peaks at $T_Q$ around 
629 days, we consistently observe a peak at approximately 
the 11-year Schwabe cycle period, which we denote by 
$T_S$. Due to the limited length of the data sets, this peak 
is very broad, so the indicated maxima should be interpreted with caution.
Another clear peak appears approximately at two-thirds of the Schwabe period, 
denoted by $T_{2/3 S}$. 
This triple harmonic of the Hale cycle, which played also a big role 
in \cite{Velasco2026}, appears to be a fairly universal 
feature of oscillatory dynamos. For example, it has also been observed 
in the Riga dynamo experiment (see, e.g., Figure 15 in
\cite{Gailitis2018}). 
Note also that the peaks in the vicinity of $T_0$ emerge as 
side peaks of its interaction with $T_{2/3 S}$.

Now focusing on the very peaks at $T_Q$, we have chosen the 
segments of the abscissa on the right-hand side panels 
(b, d and f) to be equal to those of the respective (c)-panels 
of Figures 1, 2 and 3.
When we compare those panels, we can see that the Rayleigh 
test yields slight shifts in the maxima. 
This is not surprising, given that the standard deviation 
of the values of the cosine-function in the denominator 
of Equation (2) is not included in Equations (1) and (4).

Apart from that subtlety, all peaks remain quite close to the 
theoretical QBO-period of 629.29 days. To derive the $p$-value 
for the Rayleigh power $z$, and the $N$ events, we use the 
refined expression
\begin{eqnarray}
p\approx e^{-z} \left[ 1+ \frac{2z -z^2}{4 N}-\frac{24 z-132 z^2+76z^3-9 z^4}{288 N^2}  \right] \, .
\end{eqnarray}
For the 109 merged GLE and S-flare events, we 
obtain for $z=11.47$ (measured at the optimum $T_0=633.02$\,days)
a $p$-value of $8.0\times 10^{-6}$, and $p=1.6\times 10^{-5}$ for 
the $z=10.78$ as measured 
at the theoretical QBO-period of 629.29 days.  While 
both values are approximately 
2.5 times larger than those obtained in Section 4 when 
using $\rm Corr$, they are still very small, providing again 
strong 
evidence against randomness. Thus, the Rayleigh test strengthened
the results obtained when using 
Equation (2) in the main analysis.

\section*{Appendix B}

In this Appendix, we take a closer look at the sum of the three 
waves in Equation (3) and consider the possibility that they may 
have different weights.  In fact, Figures 2-4 of 
\cite{Stefani2024} indicate that both the absolute strengths and 
the relative weights of the three tidally-triggered magneto-Rossby 
waves depend on the instantaneous value of the toroidal field. 
Let us generalize Equation (3) to
\begin{eqnarray}
s(t)&=& A \cos\left( 2\pi  \cdot \frac{t-t_{\rm VJ}}{0.5 \cdot P_{\rm VJ}}\right) +B \cos\left( 2\pi  \cdot \frac{t-t_{\rm EJ}}{0.5 \cdot P_{\rm EJ}} \right)+\cos\left( 2\pi  \cdot \frac{t-t_{\rm VE}}{0.5 \cdot P_{\rm VE}} \right) \;,
\end{eqnarray}
where A and B measure the relative weights of the first and second 
magneto-Rossby waves with respect to the third one.
While for weak toroidal fields, Figure 2 of \cite{Stefani2024}
would suggest a rather low value of $A$, the 
breakdown of the excitation of the wave with $P_{\rm VE}$ 
at the strongest fields (see Figure 4 of that paper) 
would point to a large value of $A$. 
Since correctly incorporating this subtle dependence into our 
scheme requires much more effort, here 
we can only explore some generic consequences of changing the 
relative weights.

\begin{figure}[h]
  \centering
\includegraphics[width=0.95\textwidth]{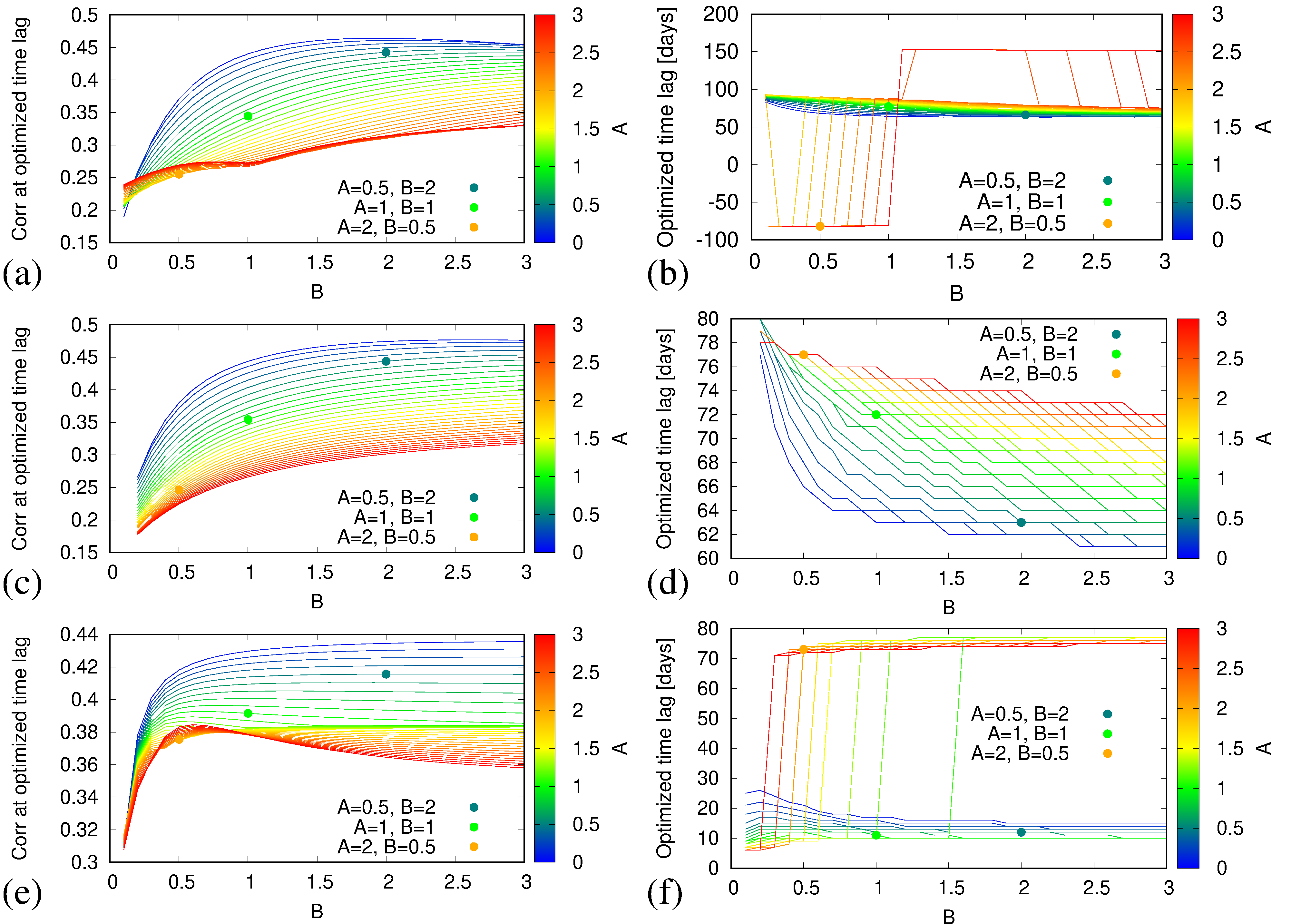}
  \caption{Dependence of the maximum correlation and the optimized time-lag on the weights $A$ and $B$ in Equation (7). Left column: Maximum correlation over $B$ for different $A$ (color coding), using
  the function $s^2(t)$ (a), 
  its time-average over 61 days (c)
  or 201 days (e). Right column: Optimal time-lag, using
  the function $s^2(t)$ (b), 
  its time-average over 61 days (d)
  or 201 days (f). The three particular combinations of $A$ and $B$, as indicated by the colored full circles, will be analyzed in more detail in the next figures.}
\end{figure}

 The left column of Figure 7 shows the maximum correlation (taken at the optimized time-lag shown in the right column) of the modified $s^2(t)$ function  with the 109 
 merged GLE and S-flare events, when $A$ and $B$ are varied. Panels (a), (c), (e) show this for $s^2(t)$ and its 
 averages over 61 and 201 days, respectively. Panels (b), (d) and (f) on the right show the corresponding optimal time lags.
 Remarkably, the correlation of 0.35 that was obtained previously for the $A=B=1$ case (green full circle in panel (a)) increases to 0.45 when we choose $A=0.5$ and $B=2$ (blue full circle). This is nearly identical to the 
 maximum correlation obtained with the optimized cosine function,
 leading hence to very similar low p-values.
 The other combination, $A=2$ and $B=0.5$, gives a reduced correlation of only 0.25, though.
These tendencies of the maximum correlation for varying $A$ and $B$ are 
quite robust when choosing longer average times (panels (c) and (e)).

The jumps of the optimal time lags, as seen in panels (b) and (f),
are  not surprising in view of the similar behaviour that was already 
visible in Figure 4(d). More surprising is  the relatively  
stable {\it positive} time-lag of about 70 days as seen in Figure 7(d).
 Why do solar events usually occur about 70 days {\it before} the maximum of the tidal forcing? One possible explanation may be related to the finite width of the dominant peaks, which is approximately 90 days according to Figures 5 and 6. Consider the possibility that the waves are predominantly excited not at the maximum of the $s^2(t)$ function, but in the time intervals of its steepest ascent or descent. Then,
 positive time lags of 50 days seem not out of reach.

\begin{figure}[h]
  \centering
\includegraphics[width=0.85\textwidth]{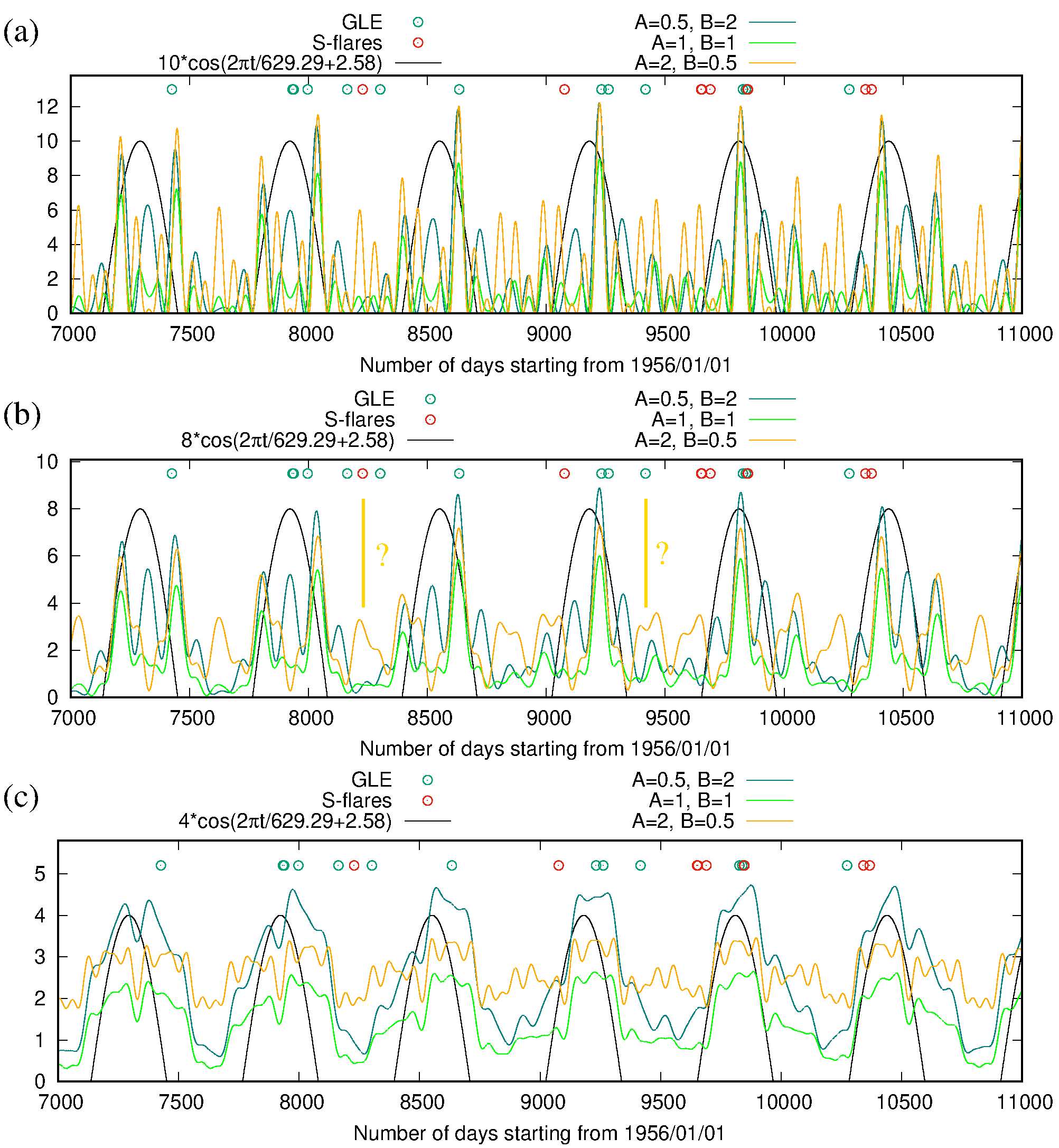}
  \caption{(a) GLE and S-flare events, the 629.29-day periodic function optimized for the merger of GLE events and S-flares, and $s^2(t)$ for three different combinations of $A$ and $B$. (b) The same as (a), but with a time average of the $s^2(t)$ functions 
  over 61 days. (c) The same as (b), but with a time average over 201 days. The gold question marks in (b) point to a possible link of solar events with side peaks of $s^2(t)$ for 
  the parameter combination $A=2$, $B=0.5$.}
\end{figure}

Figure 8 illustrates the relationship between solar events and tidal forcing over the period from 7,000 to 11,000 days after 31 December 1955. Panels (a), (b), and 
(c) show this for $s^2(t)$ and its 
averages over 61 and 201 days, respectively. Each panel 
comprises curves representing the three combinations of $A$ and $B$ that were indicated by the full circles (with the same color) in Figure 7. While the majority of solar events are concentrated around the maxima of $s^2(t)$, a few do not fit this pattern.  Interestingly, one might get the impression that they are connected to certain side peaks of $s^2(t)$, which emerge when using $A=2$ and $B=0.5$. The gold question marks in panel (b) indicate this possible link.

\begin{figure}[h]
  \centering
\includegraphics[width=0.75\textwidth]{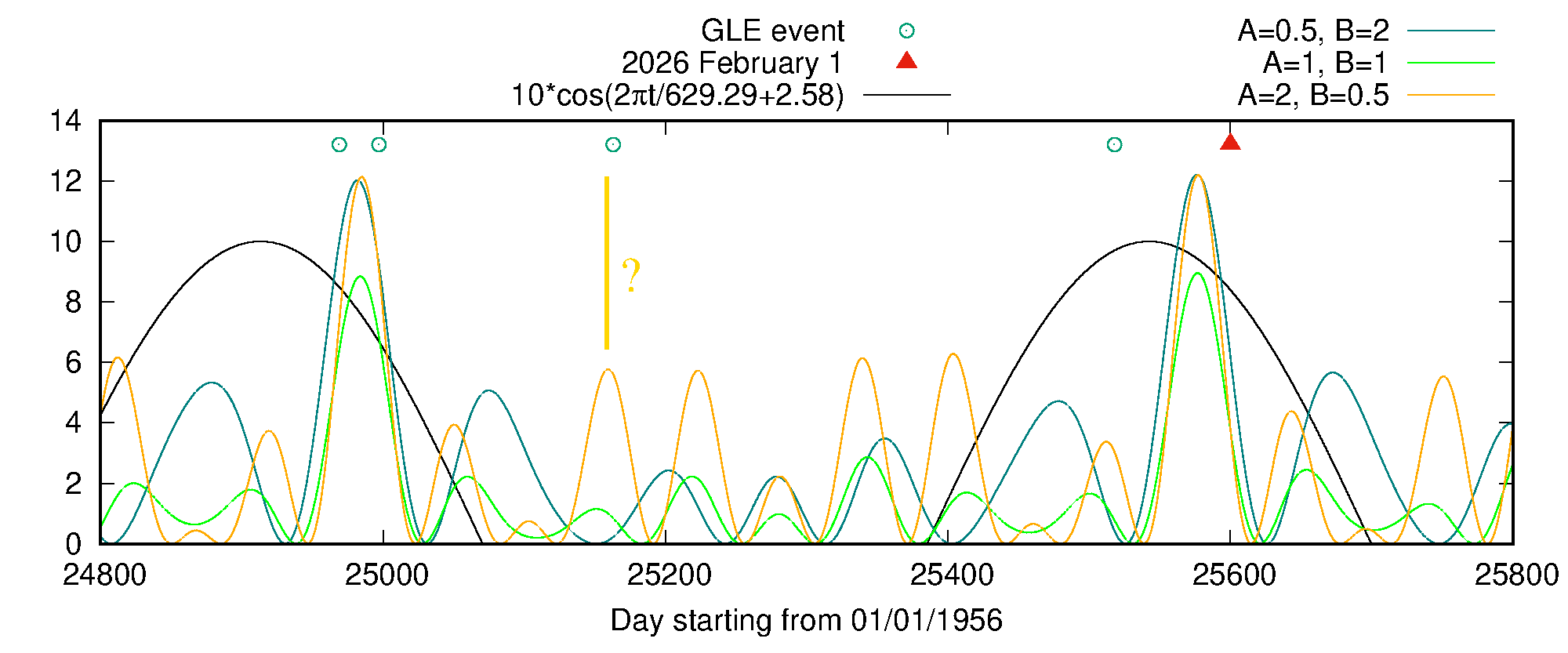}
  \caption{The last four GLE events of 2024 and 2025, together with the strong X-class flare of 2026 February 1,  the 629.29-day cosine function optimized for the merger of GLE events and S-flares, and $s^2(t)$ for three different combinations of $A$ and $B$. The gold question mark indicates the possible link
  of the third GLE event to one of the side peaks of 
  $s^2(t)$ for the parameter combination
  $A=2$ and $B=0.5$.}
\end{figure}

 A similar link might have been in effect during the last two years.
 Figure 9 is a modification of Figure 5, showing again
the last four GLE events of 2024 and 2025 and the strong X-class flare of 2026 February 1, together with the optimized 629.29-day periodic function. This time, however, we show three $s^2(t)$ functions for various combinations of A and B. While the relationship between the main peaks of the three $s^2(t)$ functions and the 2026 February 1 X-flare, as well as the first two and the last GLE events, remains unchanged, a side peak appears close to the third GLE event for $A=2$ and $B=0.5$. The gold question marks in 
panel (b) indicate a possible link between the side peaks 
of $s^2(t)$ and solar events.

In future work we will  take 
into account the dependence of $A$ and $B$ on the instantaneous toroidal field. 
In the most optimistic case, such a detailed investigation may also resolve the 
apparent contradiction that the case $A<B$ leads to higher correlation, 
while the case $A>B$ may explain the occurrence of some solar events 
between the main peaks.

\section*{Acknowledgments} 
This work received
funding from the Helmholtz Association 
in frame of the AI project GEOMAGFOR (ZT-I-PF-5-200), 
and from Deutsche Forschungsgemeinschaft (DFG) in frame of 
grant No. MA10950/1-1.
F.S. thanks Willie Soon for providing an early version of \cite{Velasco2026}, which shows evidence of a 1.7-year signal in S-flare data. He would also like to thank  
Tony Phillips for inquiring into potential impacts 
of the January 2026 alignment on solar activity.

\bibliography{glesflare}{}
\bibliographystyle{aasjournal}



\end{document}